# Two-Photon-Induced Direct 3D Printing of Freeform High-Index Phase-Change $Sb_2S_3$ Nanostructures

*Abhrodeep Dey[1,2], Andrea Dellith[1], Anne Sauer[3], Uwe Hübner[1], Henrik Schneidewind[1], Markus A Schmidt[1,4,5], Astrid Bingel[3], Volker Deckert[1,2,4,6], Jer-Shing Huang[1,2,4,7,8*], Wei Wang[1,2*]*

E-Mail: jershing.huang@ipht-jena.de; w.wang@uni-jena.de

[1] Leibniz Institute of Photonic Technology (IPHT), 07745 Jena, Germany.

[2] Institute of Physical Chemistry, Friedrich Schiller University Jena (FSU), 07743 Jena, Germany.

[3] Functional Optical Surfaces and Coatings, Fraunhofer IOF, 07745 Jena, Germany

[4] Abbe Center of Photonics and Faculty of Physics, FSU, 07745 Jena, Germany.

[5] Otto Schott Institute of Material Research, FSU, 07745 Jena, Germany.

[6] Jena Center for Soft Matter (JCSM), FSU, 07743 Jena, Germany.

[7] Research Center for Applied Sciences, Academia Sinica, 11529 Taipei, Taiwan.

[8] Department of Electrophysics, National Yang Ming Chiao Tung University, 30010 Hsinchu, Taiwan.

Abstract: Chalcogenides have recently emerged as an important class of phase-change materials (PCMs) for nanophotonics, owing to their very high refractive index (*RI*) and low optical loss in the visible to near-infrared range. They exhibit an ultralarge *RI* change (> 0.7) upon phase transition, which can be triggered by multiple stimuli such as electrical bias, laser illumination or thermal heating. These properties make them highly appealing materials for flat optics and metasurface applications. Current nanophotonic implementations of chalcogenide PCMs mostly rely on two-dimensional (2D) or quasi three-dimensional (3D) thin-film patterning based on the coating of chalcogenide materials from a solid-state target. This limits fast prototyping of 3D freeform micro- and nanostructures, thus restricting geometric design freedom and device functionality. Here, we demonstrate a solution-phase direct printing of chalcogenide PCMs into functional structures. The method is based on dip-in two-photon-induced solidification (DITPS) of a specially synthesized antimony trisulfide ($Sb_2S_3$) precursor solution. Direct printing with DITPS is simple, maskless, fast and cost-effective, enabling true freeform 3D printing of photonic devices with sub-micron resolution. We show direct writing of $Sb_2S_3$ helices with different wire cross-section profiles on gold and ITO substrates, as well as functional planar Fresnel zone plates (FZPs) and computer-generated hologram metasurfaces (CGHMs) in a single printing step. This freeform DITPS approach thus enables rapid 3D prototyping of high-index metasurfaces and opens a route to integrating high-index PCMs into existing photonic architectures and device platforms.

## 1. Introduction

The emergence of metasurfaces designed to precisely manipulate electromagnetic waves has transcended the limitations posed by traditional bulk optics.[1, 2] Leveraging on planar arrays of flat optical components for imaging and sensing,[3, 4] polarization control devices,[5] plasmonic sensors,[6, 7] dynamic holographic displays,[8] augmented and virtual reality (AR/VR) displays[9] and high-Q resonant structures based on bound states in the continuum[10] has positioned metasurfaces as a cornerstone technology for next generation photonic systems.[11, 12] Metasurfaces fundamentally consist of metal or dielectric structures called meta-atoms arranged in a periodic or aperiodic configuration. However, dielectric meta-atoms support high-efficiency wavefront manipulation with minimal optical losses, combining strong Mie-type field confinement and low absorption for superior performance over its metallic counterpart.[13] In particular, integrating PCMs with high *RI* into dielectric metasurfaces introduces dynamic, non-volatile tuning of the optical response, surpassing the static functionality of conventional passive designs. Moreover, such high-index PCM platforms enable active modulation of electric and magnetic Mie resonances in subwavelength meta-atoms, providing highly efficient wavefront control in compact flat-optical devices.[14] $Sb_2S_3$ has recently reemerged as a high *RI*, ultralow-loss ($k < 10^5$ at $\lambda = 1550$ nm), wide bandgap (1.7–2.0 eV) material.[15] $Sb_2S_3$ can also transition between two distinct material states: amorphous ($a$-$Sb_2S_3$) and crystalline ($c$-$Sb_2S_3$) while displaying significant index contrast ($\Delta RI = 0.6$).[15] It also has ultralow loss in visible and infrared spectral regime, which is over two orders of magnitude better than traditional PCM, such as Germanium Selenium Telluride alloy enabling application in programmable integrated photonic circuits, switchable metasurfaces and diffractive optical elements.

Traditional nanofabrication approaches typically follow thin film deposition of $Sb_2S_3$ via sputtering, evaporation, or pulsed laser deposition, followed by pattern definition using electron beam lithography (EBL)[16, 17] or photolithography[18] and subsequent pattern transfer by reactive ion etching or ion-beam etching. While these top-down workflows enable precise control over meta-atom geometry and optical response, they typically involve multi-step repeated cycles of deposition, lithography and etching which increase process complexity.[19] It also constrains the attainable 3D design space and are often limited by slow throughput, complex workflow and expensive setups.[20]

Two-photon absorption lithography (TPAL) has emerged as a powerful direct-write additive manufacturing technique that enables free-form, 3D nanostructuring with sub-micrometer feature sizes, high design flexibility and relatively simple, maskless workflows compared to conventional multi-step lithography. Recent publications explored commercially available resist IP-Dip,[21, 22] and materials such as, hydrogen silsesquioxane (HSQ)[23] and Silica glass[24, 25] to print 2D and 3D structures using TPAL. However, since standard TPAL relies on photo-crosslinking of polymeric resists, its use is largely restricted to low *RI* polymer structures, making it challenging to directly realize high *RI* dielectric or phase-change meta-atoms required for low-loss, high-efficiency metasurfaces. Low-index ($RI < 1.7$) materials often demand very high-aspect ratio structures[26] to sustain efficient phase and amplitude control.[27] To achieve more efficient light modulation enabling performance comparable to EBL fabricated dielectric metasurfaces, high *RI* materials ($RI > 1.7$) are required.[28, 29] High *RI* materials allow further miniaturization of meta-atoms while also enabling exploitation of the out-of-plane dimension via printing in an additional spatial degree of freedom.[30] High *RI* materials also enable metasurfaces to function in solution-based environments that demand a significant *RI* contrast. Recent advances, such as the demonstration of TPAL printed $TiO_2$ 3D photonic crystals with a complete bandgap in the visible using a customized titanium ion-doped resin, have pushed the accessible *RI* into the 2.4 – 2.6 range.[31] But this still falls short of the freeform 3D printing of high *RI* platforms desired for many dielectric and phase-change metasurface applications. Additionally, the capability to 3D print freeform high *RI* metastructures enables complex chiral architectures such as helices and twisted networks, while markedly enhancing their chiroptical response compared to low-index polymer or silica.[7, 32] Freeform 3D architectures also break out-of-plane symmetry unlocking full vectorial control over optical responses inaccessible by planar 2D structures. Moreover, in TPAL the voxel-based exposure confines absorption to the focal volume of an otherwise transparent resin, preventing line-of-sight curing and avoiding stair-stepping effect. This enables voxel-confined writing inside bulk volumes and on pre-existing structures (e.g., fiber end facets[33] or microfluidic channels[34]). Therefore, the key to using 3D direct printing for high-index chalcogenide PCMs is the availability of a precursor solution that can be solidified under two-photon laser irradiation.

In this work, we report the first experimental demonstration of a precursor solution-based DITPS route to fabricate complex, maskless, true freeform rapid-prototyping of 3D *a*-$Sb_2S_3$ structures. Using tightly focused 780 nm fs-laser pulses, DITPS is locally triggered within the precursor solution, yielding spatially resolved conversion of the solution to *a*-$Sb_2S_3$. Building on this concept, we establish a process window and development protocol that enables reproducible fabrication of complex 3D freeform *a*-$Sb_2S_3$ structures with controlled geometry and stable optical response. We study the impact of focal spot shape and demonstrate freeform 3D helices with two different wire cross-section on Au- and ITO-coated $SiO_2$ substrates. The fabricated structures were systematically characterized using a range of optical and analytical techniques. Finally, we fabricate quasi-3D FZPs and CGHM structures as representative optical application.

## 2. Results

### 2.1 DITPS of precursor solution

To initiate DITPS process for 3D printing, the precursor solution must meet specific criteria.[35] First, it should exhibit high transmittance at the laser wavelength to avoid any unwanted thermal effect due to one photon absorption and to ensure high field intensity at the focal spot. Second, it should possess a large two-photon absorption cross-section such that two-photon-induced solidification (TPS) happens effectively at the focal spot. Finally, the solidified structure must become insoluble and preferably firmly attached to the substrate to ensure safe extraction of the printed structures during development.

In this work, the precursor of $Sb_2S_3$ to be printed is obtained by the reaction of $Sb_2O_3$ in a butyldithiocarbamic acid (BDCA) solution. BDCA has been used previously to dissolve metal oxides ($M_xO_y$) and hydroxides ($M(OH)_x$) to form their respective metal-dithiocarbamate complexes,[36] which can be attributed to the highly active thiol-amine group in BDCA. Therefore, following reported protocols, the BDCA solution was synthesized by the reaction of $CS_2$ and n-butylamine.[37] Next, $Sb_2O_3$ was dispersed in the BDCA solution to prepare the antimony butyldithiocarbamic acid (Sb-BDCA) precursor solution (synthesis details in Methods section). Upon absorption of two photons at 780 nm, Sb-BDCA complex in the solution decomposed to generate $Sb_2S_3$ in its amorphous, solidified form. This is evident from previous studies that investigated the thermochemistry of metal-dithiocarbamate complexes to explain the formation mechanism of metal sulfides during thermal decomposition.[38, 39]

The precursor solution can be utilized in both positive-tone and negative-tone fabrication approach. Very recently our work demonstrated BDCA as an organic etchant to selectively pattern *a*-$Sb_2S_3$ 2D and 2.5D micro/nanostructures via positive-tone (reductive patterning) laser lithography. [40] The negative-tone patterning (additive) can be achieved via various lithography methods, including electron beam lithography (EBL)[41, 42] ultraviolet photolithography, and thermal scanning probe lithography and TPAL.[43] A detailed comparison between the g-EBL, positive-tone and negative-tone laser direct writing of *a*-$Sb_2S_3$ structures via the Sb-BDCA route is included in Table S1 in the Supplementary information. Previous work has shown that TPAL of Sb-BDCA enables the fabrication of 3D double layer honeycomb photonic structures.[43] However, these demonstrations did not provide free-standing and 3D freeform *a*-$Sb_2S_3$ structures that withstood development.

Figure 1 shows the general process workflow to obtain *a*-$Sb_2S_3$ structures via DITPS using a commercially available two-photon printer (GT2, Nanoscribe). Initially the precursor solution was drop-casted on the Au or ITO coated $SiO_2$ substrate. Following this the objective lens (63x, NA = 1.4) was "dipped in" to the precursor solution. The solution also acts as an immersion medium for the objective lens as its refractive index ($n$ = 1.5749) closely matches that of the lens immersion oil ($n$ = 1.518). DITPS was conducted by moving a tightly focused 780 nm fs-laser inside the solution along a pre-designed path guided by galvo mirrors with pulsed mode. Finally, the excess solution was washed away with toluene to reveal the laser exposed region.

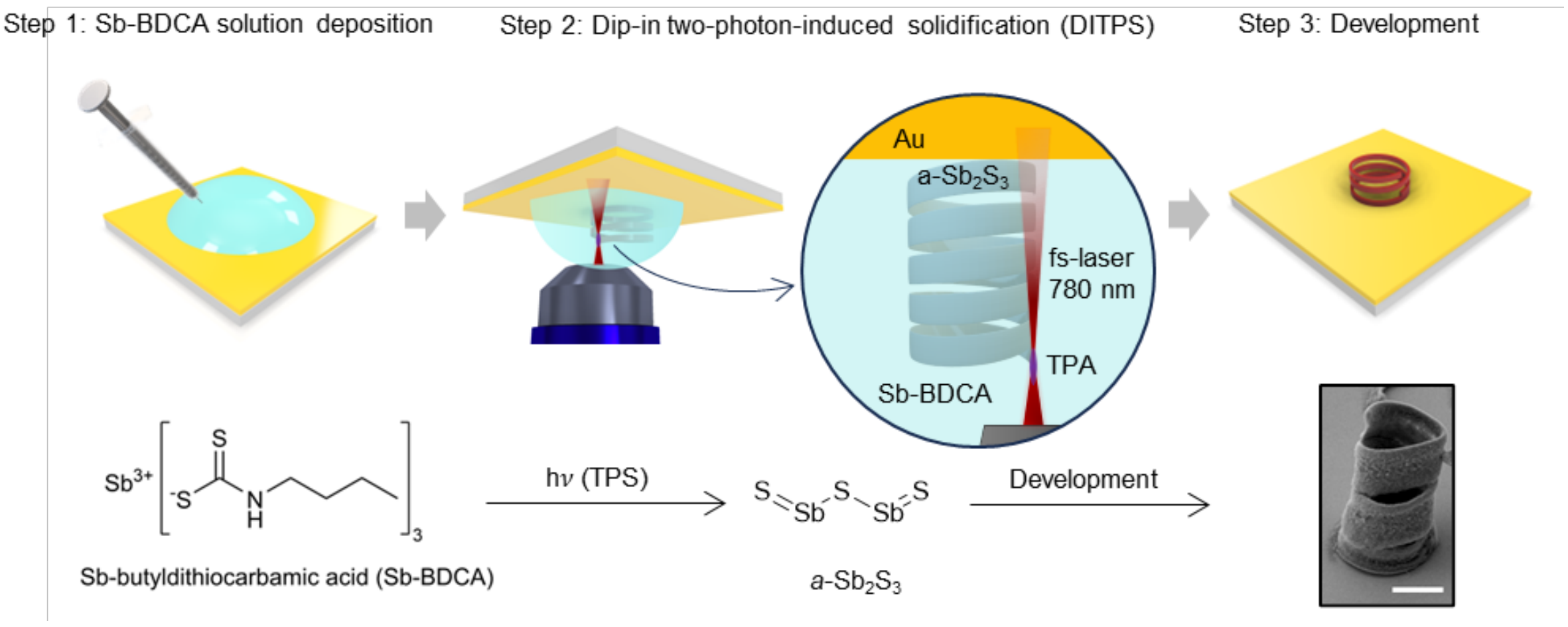

**Figure 1.** Schematic of fabrication workflow to obtain 2D and 3D freeform *a*-$Sb_2S_3$ structures via DITPS on Au-coated $SiO_2$ substrate. Similar workflow was followed for ITO-coated $SiO_2$ substrate. SEM 45° tilted view of single helix structures. Scale bar: 5 μm.

The transmittance spectrum of the Sb-BDCA solution for the spectral range between 350-800 nm is shown in Figure 2a. As seen, the solution has very high transmission at 780 nm and rather low transmission at 390 nm (two-photon wavelength), suggesting the suitability of the solution for TPS. The transmission spectrum of *a*-$Sb_2S_3$ is also provided in Figure 2a for comparison. The gradual increase in transmittance with increasing wavelength for a thin film of *a*-$Sb_2S_3$ agrees well with previously reported results.[44] The TPS rate scale with the square of the intensity ($I^2$) and the number of molecules present. Therefore, it increases as the beam area is reduced at the focus. TPS initiation requires a sufficiently high photon density, which is achieved in the tightly focused laser focal volume inside the resist. For effective excitation, the sum of the energies of the two coherently absorbed photons must be approximately equal to or slightly larger than the electronic transition energy gap. Therefore, knowing the intensity profile of the focal spot is important for estimating the area of effective TPS in the Sb-BDCA solution, i.e., the single exposed voxel (volume pixel) area.

Before fabricating $Sb_2S_3$ structures, it is also necessary to observe how the voxel shape varies with and without a substrate. This is particularly important when the substrate is a reflective metallic surface.[41] Hence, numerical simulations using Finite-Difference Time-Domain (FDTD) method were performed to simulate the shape of a single voxel. The simulation was carried out in two different environments, namely in a homogeneous medium of only Sb-BDCA solution (without any substrates) and with multi-layered $SiO_2$/Au/Sb-BDCA solution interface (simulation details provided in Methods section). Left panel of Figure 2b shows the simulated cross-sectional profile of $I^2$ (where $I = |E|^2$) at the focal spot of a tightly focused (NA = 1.4) x-polarized

Gaussian beam in homogeneous Sb-BDCA environment. The source beam is centered at 780 nm impinges from below (indicated by green dashed arrows). The simulated beam profile shows a typical focal spot in a homogeneous medium. Upon focusing the beam onto the surface of an Au film (thickness = 50 nm) on $SiO_2$ substrate immersed in precursor solution, the focal spot revealed nodes and anti-nodes due to the interference of the incoming and reflected light (center panel, Figure 2b). It was observed that by shifting the focus 1 µm downwards, i.e., away from the Au/Sb-BDCA interface towards -Z direction, the intensity maximum reduced in magnitude but the interference pattern remained. The simulation results thus verified three key findings. Firstly, the interaction between the monochromatic source waves and their reflection at the Au/Sb-BDCA boundary gives rise to spatially fixed-phase standing waves.[45] This manifests as highly localized high-intensity antinodes at the interference maxima of the standing-wave field, consistent with our previous report. [41] However, the high-NA focus generates a continuum of angular components. So, the reflected field does not form a strictly 1D standing wave and the calculated intensity minima are not exact zeros, especially away from the optical axis. As a result, the experimentally fabricated structures show only weak axial modulation and remain continuous, even though the underlying optical field exhibits distinct antinode and node regions. Secondly, the focal spot intensity at the anti-nodes decreases as the focal spot gradually shifts away from the Au/Sb-BDCA interface due to reduced interference. This indicates that the laser beam fluence must be controlled accordingly to obtain a uniformly printed profile. Thirdly, the voxel exhibits the well-known ellipsoidal shape with a high axial-to-lateral aspect ratio, as expected for a tightly focused high-NA beam in two-photon absorption. Interestingly, the experimentally fabricated structures showed the lateral resolution of a single voxel as ~280 nm, with an axial resolution of ~3.7 µm. This relatively higher axial-to-lateral aspect ratio of the voxel can be attributed to the high laser power (~67.5 mW) dose, which gives rise to "focal spot duplication" and "radical diffusion-dominated voxel growth" as observed by Sun et al.,[46] where the voxel shape directly follows the power-dependent elongation of the squared focal intensity distribution ($I^2$) near threshold.[47] Moreover, TPS proceeds relatively faster in the high-intensity focal spot isophotes at the anti-node voxel regions (middle panel, Figure 2b) giving rise to solidified growth of *a*-$Sb_2S_3$ globules with gradually increasing degree of solidification. Additionally, due to the high-index and low absorption of *a*-$Sb_2S_3$ at incident wavelength of 780 nm, intensity hot-spots emerge at the nodal positions between the already formed *a*-$Sb_2S_3$ globules, where TPS occurs forming an elongated chain of interconnected voxels. This can be modelled in the FDTD environment by placing *a*-$Sb_2S_3$ spheres with diameter similar to the lateral resolution of a single voxel at the anti-node intensity maxima position close to the Au/Sb-BDCA interface (right panel, Figure 2b). The simulated results agreed well with the experimentally printed structure on 50 nm Au-coated $SiO_2$ substrate (Figure 2c, single coil helix) where a high aspect ratio topography can be observed. Additional SEM images of freeform 3D structures fabricated both on Au- and ITO-coated $SiO_2$ substrate is included in Figure S2 in the Supplementary Information. Given the elongated nature of the focal spot, two different types of helices with two different wire cross sections, i.e., parallel or perpendicular to the helix axis were designed and fabricated (Figure 2c). The base radius was varied between 2.5, 3 and 3.5 µm for both helices. For the helix with axis pointing out-of-plane, perpendicular to the substrate, the pitch was set at 2.5 µm. Whereas for the other case, the pitch was 3.0 µm. The anisotropic focal spot profile provided an additional degree of freedom for the design of the structures.

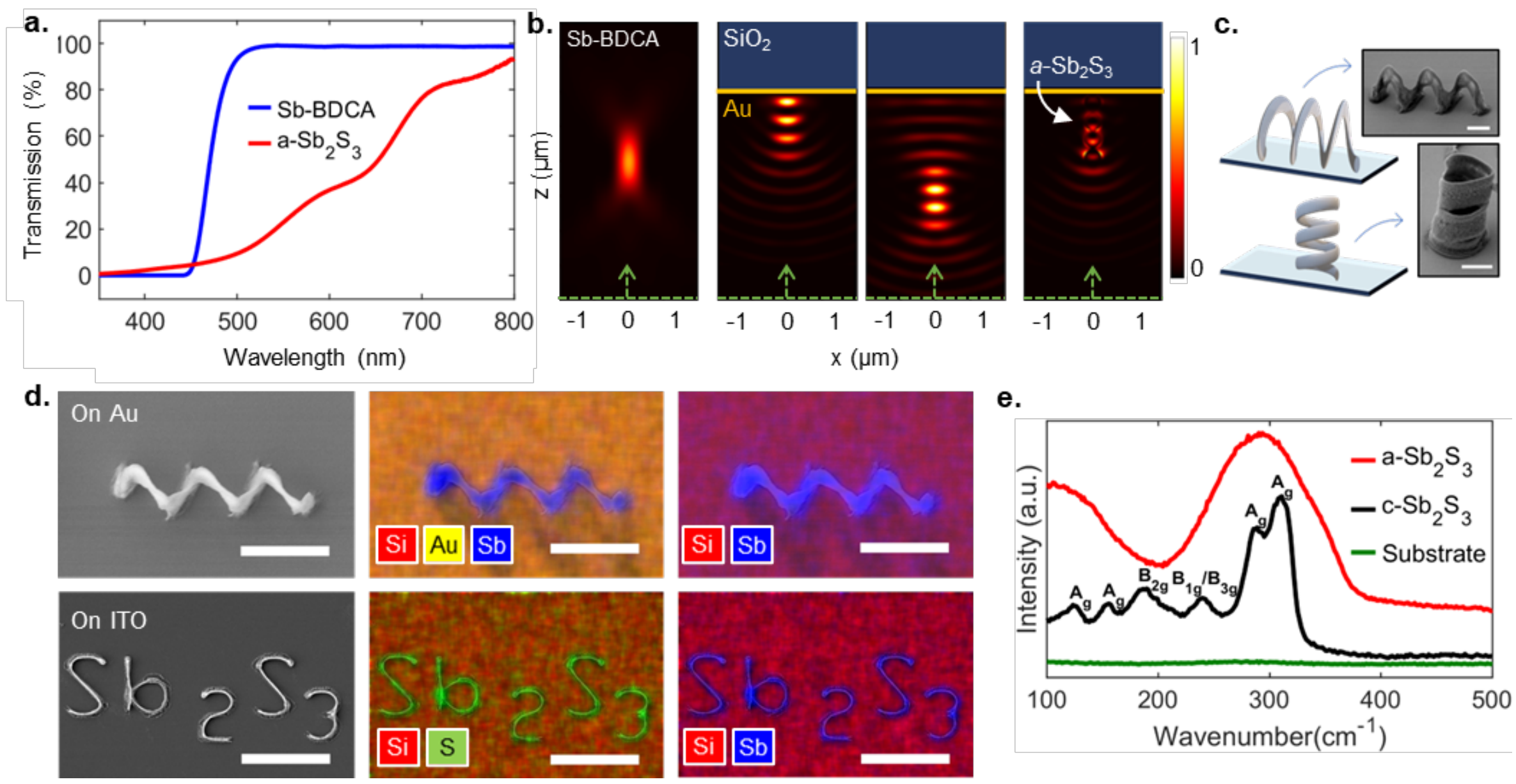

**Figure 2.** Optical response and structural characterization of Sb–BDCA. **(a)** Transmission spectra of a pristine Sb-BDCA solution (blue) and a thin film of *a*-$Sb_2S_3$ with a thickness of 140 nm (red). **(b)** Simulated cross sectional profile of squared intensity ($I^2$), representing the effective field distribution of the focal spot of a focused source illuminating from the bottom for TPS process. A tightly focused Gaussian light source traverses from below (green dashed arrow pointing towards propagation direction) towards (left panel) pure Sb-BDCA, (middle panel) multi-layer $SiO_2$/Au/Sb-BDCA solution interface where the source is placed -2.5 and -3.5 µm away from the Au/Sb-BDCA interface towards –Z direction, and (right panel) *a*-$Sb_2S_3$ globules placed at the intensity maxima with the source placed -2.5 µm away from the Au/Sb-BDCA interface. **(c)** 3D illustration and 45° tilted SEM images of vertically and horizontally oriented 3D helices, fabricated with anisotropic focal spot profile. Scale bar: 5 µm **(d)** (Top panel, left) SEM image and (middle and right) EDX maps of a 3D horizontal helix structure printed on 50-nm Au film on $SiO_2$ substrate. The EDX elemental maps confirm the presence of antimony (Sb) from the printed pattern, gold (Au) and silicon (Si) from the surrounding unexposed area. Presence of sulfur (S) could not be mapped due to the strong overlap of S-Kα-line with the Au-Mα-line. Further details are provided in Figure S3 in the Supplementary information. (Bottom panel, left) SEM image and (middle and right) EDX elemental maps of the letter „$Sb_2S_3$“ printed on a 70-nm thick ITO layer on a $SiO_2$ substrate. The EDX elemental maps confirm the presence of S in the printed region and Si and Sb from the unexposed surrounding areas. Scale bar: 10 μm. **(e)** Raman spectra of *a*-$Sb_2S_3$ (red), *c*-$Sb_2S_3$ (black) and ITO-coated $SiO_2$ substrate (green).

Having discussed the focal spot profile, we further analyze the composition of the printed structures. Elemental mapping via Energy-dispersive X-ray (EDX) spectroscopy was performed to analyze the elemental composition of the developed structures both on Au- and ITO-coated $SiO_2$ substrate. The analysis was performed on both 2D and 3D structures to validate the formation of *a*-$Sb_2S_3$. Figure 2d (top panel) show the SEM image and the corresponding EDX maps of a freeform 3D helix structure printed on the surface of a 50 nm gold film coated on a $SiO_2$ substrate. The bottom panel in the same figure show the SEM image and EDX elemental maps of the 2D "$Sb_2S_3$" printed on a 70 nm ITO layer coated on a $SiO_2$ substrate. For the helix, a 45° tilted SEM image is shown in the inset. For the 2D and 3D structures printed on both substrates, EDX maps confirm the presence of Sb and S in the laser exposed regions, with strong signals overlaying the printed structures. Raman spectra were recorded for the printed *a*-$Sb_2S_3$, the thermally annealed *c*-$Sb_2S_3$ and ITO-coated $SiO_2$ substrate to confirm the phase-change characteristics of the structure (Figure 2e). [41, 48]

## 2.2 Binary Fresnel zone plates and metasurface-based hologram

To demonstrate the high-index optical device printing capability of the reported fabrication technique in optics, binary FZPs were printed on the 50 nm Au-coated and 70 nm ITO-coated $SiO_2$ substrates. To compare the fabrication quality on both substrate platforms, FZPs with same geometry, i.e., same number of concentric rings and 400 μm focal length were printed. The FZPs were designed based on the equation: $r_n = \sqrt{((n^2\lambda^2/4) + nf\lambda)}$, where $r_n$ is the radius of the $n^{th}$ zone, the operational wavelength $\lambda$ is 633 nm, the designed focal length $f$ is 400 μm and $n$ is the index of $n^{th}$ zone with a maximum number of 25. Figure 3a shows the optical setup used to acquire the image of a USAF 1951 target using the printed FZPs. The USAF 1951 test target is ideal to determine imaging resolution of the lenses, because its line pairs cover a defined range of feature sizes and gaps that span the expected resolution of the device, enabling a direct determination of the smallest resolvable structures. Figure 3b shows that the higher resolution elements of USAF target are placed near the center, while the low-resolution elements are pushed towards the periphery of the target. One line pair (lp) corresponds to one black bar and one adjacent white space. The group and element indices together uniquely define vertical and horizontal resolution at discrete spatial frequencies in the object plane. To determine the resolution ($R$) in lp/mm, the following equation is used: $R = 2^{[K+((N+1)/6]}$, where $N$ is the element number and $K$ is group number. We observed that utilizing FZPs printed both on Au and ITO substrates, the $0^{th}$ group $6^{th}$ element could be imaged. This is evident from Figure 3c top and bottom panel. Therefore, based on the resolution ($R$) formula, the element 4, 5 and 6 in group 0 have a resolution of 1.41, 1.58 and 1.78 lp/mm under broadband white light source.

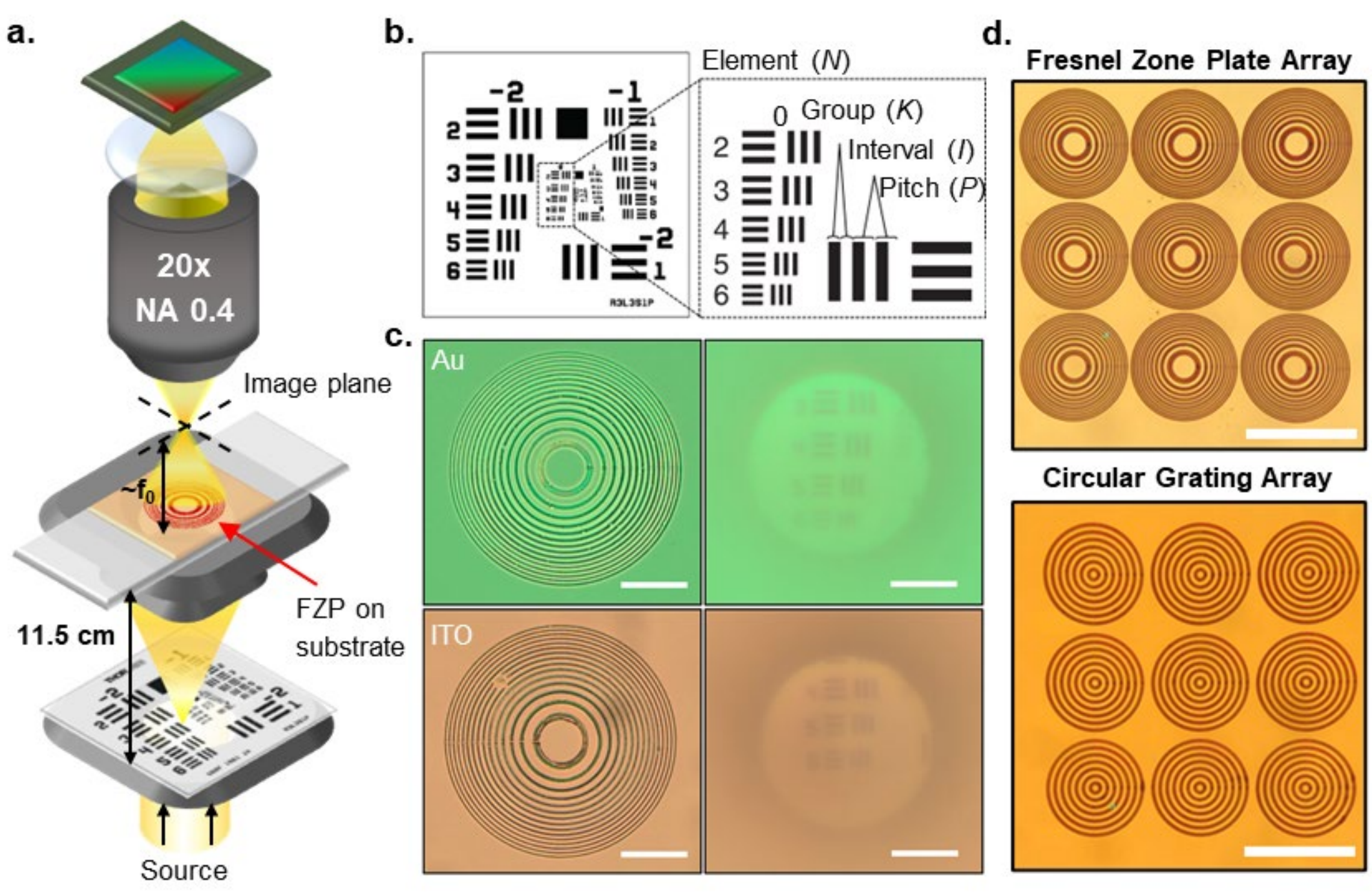

**Figure 3.** Imaging of USAF 1951 target with $a$-$Sb_2S_3$ FZPs printed on Au and ITO substrates by DITPS. **(a)** Illustration of optical setup configuration used to image the USAF target. The test chart is placed 11.5 cm away from the FZP (focal length ($f_0$) = 400 µm). The intermediate image forms at a distance around $1f_0$ distance behind the FZP which is captured by an 20x objective lens with NA = 0.4. **(b)** USAF 1951 target pattern layout using Group 0 line pair (lp) specification for imaging. **(c)** Bright field transmission image of 400 µm FZP imaged with 20x, NA = 0.4 objective lens on (top left) Au and (bottom left) ITO substrate. Group 0 line pairs imaged show well resolved line pairs until element 6 for both (top right) Au and (bottom right) ITO substrates. Scale bar: 25 µm **(d)** (Top panel) FZP array with focal length = 500 µm and (bottom panel) circular grating array. Scale bar: 150 µm.

Figure 3d depicts arrays of FZP (top panel) and circular gratings (bottom panel) printed on an 18 x 18 $mm^2$ substrate. In order to transition from single devices to practical large-area applications, 3 x 3 array of FZPs with focal length = 500 µm designed for an operating wavelength $\lambda$ = 532 nm were printed (Figure 3d, lower left panel). Furthermore, 3 x 3 array of concentric circular gratings resembling a Bull's eye aperture were printed (Figure 3d, lower right panel) to probe patterning quality over the Galvo scan field, confirming consistent performance for different ring radii. These sequentially printed arrays are essential to reveal fabrication uniformity, repeatability and validate large-scale printing capability.

To further demonstrate the ability to fabricate densely encoded, maskless and complex holographic phase patterns, a CGHM was fabricated using the DITPS method. Holographic metasurfaces offer an efficient and compact alternative to bulky optical components by spatially encoding varying phase profiles within the metastructure. This is especially useful in miniaturization of optical systems for applications in encryption,[49] imaging[50] and displays.[8] To generate the hologram, numerical calculations with following steps were utilized: (i) calculating the Fourier transform of the transmission function of the object $t(x,y)$, (ii) simulation of $S(u,v) = exp\ (-i\pi\lambda z(u^2+v^2))$ where (u,v) denote Fourier domain coordinates, (iii) multiplication of t(x,y) and S(u,v) and (iv) calculating the inverse Fourier transform of the result and obtaining the square of the absolute value of the result from previous step.[51] The obtained binary version of the phase mask (*S(u,v)*) is printed via DITPS on the substrate to form a CGHM-based phase mask (Figure 4a, left panel). The printed CGHM is then imaged via the optical setup (Figure 4a, right panel and Figure 4b) to form the reconstructed image.

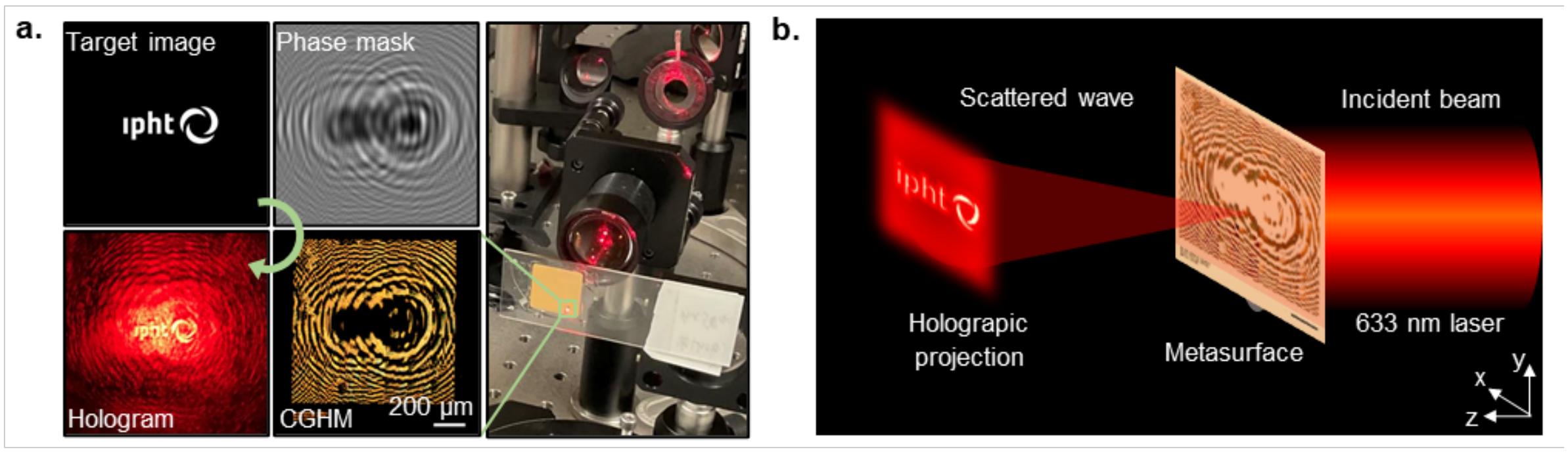

**Figure 4.** DITPS-based printing and imaging of holographic metasurface. **(a)** (Left panel) Flowchart of design, print and imaging of the letters "IPHT" via CGHM. (Right panel) The optical setup used to image the CGHM. **(b)** Schematic illustration of the holographic projection mechanism.

## 3. Discussion

In this work we demonstrated a robust solution-phase DITPS technique for fabricating monolithic planar 2D and 3D freeform *a*-$Sb_2S_3$ structures using a Sb-BDCA precursor solution. Our approach enables the freeform 3D printing of a high-index, low loss, phase-change chalcogenide achievable through a maskless, direct dip-in, two-photon induced solidification process that was previously not demonstrated. The thermal decomposition of the precursor solution under fs-laser, induced two-photon solidification that led to the formation of *a*-$Sb_2S_3$.[39, 40] Moreover, our solution-phase fabrication approach greatly simplifies and presents an alternative route to the existing multiple deposition, lithography, etching and extensive post-processing for fabrication of high-index chalcogenides, thus making the DITPS process efficient and cost-effective. While beyond the scope of this work, the same *a*-$Sb_2S_3$ structures can be reversibly tuned between its crystalline and amorphous state through optical, thermal or electrical stimulus.[15, 52, 53] This enables the fabrication of dynamically tunable structures which extend its application to reconfigurable metasurfaces in sensing, imaging, data storage and telecommunication.[54] In order to unlock the full potential of DITPS, we propose several research directions that can be undertaken. Firstly, since BDCA can be used to dissolve various types of metal hydroxides and oxides. Therefore, we envision DITPS may be extended to other metal chalcogenide systems, opening a route to 3D print other materials such as ZnS, PbS, $Sb_2S_{3-x}S_{ex}$. Secondly, DITPS can also be combined either with multi-beam, spatiotemporal focusing femtosecond projection TPA lithography (FP-TPL)[55] or with large-area meta-lens generated focal spot array to parallelize and produce centimeter scale structures.[56] Lastly, in order to fabricate structures with sub-wavelength feature size, we visualize merging our fabrication approach to novel techniques such as stimulation emission depletion (STED) lithography[57] or light matter co-confined multi-photon lithography.[58] This recently reported technique utilizes concepts similar to using doughnut-shaped STED like inhibition beam to deplete excited states around the voxel periphery.

## 4. Methods

Materials: Antimony (III) oxide ($Sb_2O_3$, 99.99%) was obtained from Thermo Fischer Scientific, 1-butylamine (($CH_3(CH_2)_3NH_2$), 99.5%) and carbon disulfide ($CS_2$, 99.9%) was purchased from Sigma-Aldrich, and ethanol ($CH_3CH_2OH$, 99.99%) was acquired from Merch KGaA. For the gold coated substrate, 50 nm of gold was sputtered onto 1 mm thick fused silica ($SiO_2$) substrate and for the ITO coated substrate, 70 nm thick

ITO layer was sputtered on 1 mm thick $SiO_2$ substrate with lateral dimension of 18 x 18 $mm^2$.

Sb-BDCA synthesis: For a typical Sb-BDCA precursor solution, 1.0 mmol of $Sb_2O_3$, 2.0 mL of ethanol, and 1.5 mL of $CS_2$ were added to a 50 mL three-neck flask and stirred magnetically at room temperature. Then, 2 mL of 1-butylamine was rapidly injected into the mixture and the reaction was allowed to proceed under continuous stirring for 17.5 hours, yielding a clear and uniform solution. After exposure to air, the resist should rest for 30 minutes prior to the printing process. The Sb-BDCA solution can be handled in air for up to 12 hours. Prolonged exposure may hinder the development process.

Dip-In two-photon-induced solidification (DITPS): All structures were designed using the Describe programming language (Nanoscribe GmbH) and patterned with a Nanoscribe Photonic Professional GT2 system at room temperature. A 63x, NA = 1.4 objective lens with dip-in laser lithography configuration was directly immersed in the Sb-BDCA precursor solution drop-casted on respective substrates. The exposure time used was 3.2 ms with peak laser power 67.5 mW. Following the printing process, the substrates were subjected to development, during which residual Sb-BDCA was removed using toluene, followed by a brief rinse with Novec™ solution (3M) and subsequent air drying.

Device and material characterization: All energy dispersive X-ray (EDX) analyses were done using a state of the art 30 $mm^2$ silicon drift detector (SDD) by BRUKER (BRUKER Nano GmbH, Berlin, Germany) and the Esprit spectra evaluation software package. The specified energy resolution of the detector at 5.9 keV (Mn-Kα) is 129 eV.

Numerical simulations: The electric field intensity distribution of the Sb-BDCA structure on a gold substrate was investigated using commercially available finite-difference time-domain (FDTD) method simulation package (Finite Difference IDE version 8.31.3683, Ansys Lumerical Software ULC, Canada). Three separate simulation models were prepared: first, consisting solely of pure Sb-BDCA (refractive index *RI* = 1.57) second, comprising a top Sb-BDCA layer (*RI* = 1.57) followed by a 50 nm thick Au film atop an $SiO_2$ substrate and third, four spherical *a*-$Sb_2S_3$ structures positioned at the anti-node intensity maxima extracted from the $SiO_2$/Au/Sb-BDCA simulation. The excitation source was defined as an X-polarized, converging Gaussian beam centered at a wavelength of 780 nm, with a numerical aperture NA of 1.4, propagating along the +Z direction. This source configuration was consistently used for all simulations in this study. 2D FDTD simulation region with perfectly matched layer (PML) boundary condition were applied along X and Z axes respectively to absorb outgoing electromagnetic waves and prevent reflections. A 2D frequency and power monitor was positioned along the XZ plane to visualize the electric field intensity ($I = |E|^2$) distribution across the structure. The beam cross-sections were then calculated and plotted separately by squaring the extracted intensity ($I^2$) from the 2D frequency and power monitor.

## Author Contributions

WW and JSH conceived the idea. AD and WW synthesized the material and designed the experiment. AD and WW performed the dip-in two-photon-induced solidification (DITPS). AD, AS and AB performed optical inspection/characterization. AD and WW performed the FDTD simulation. UH and HS prepared the substrates. AD (Andrea Dellith) performed the AFM, SEM and EDX characterization. MAS and WW were involved in the design of the FZPs. AD wrote the first draft of the manuscript. All authors contributed to the writing of the manuscript. JSH and VD supervised the study.

## Acknowledgements

WW, VD, JSH, and AD thank for financial support from The DFG via SFB 1375 NOA (Project No.: 398816777). JSH and AD appreciates support via the DAAD-PPP-Japan Tsukuba University (project No.: 57710865), IRTG 2675 Meta-Active (Project No.: 437527638). WW thanks for financial support through the IPHT Innovation Project 3D-HiRes 2021/2022 (Project No.: K690082) and the DFG via CRC-TRR 234 CataLight (Project No. 364549901). This work is further supported by the BMBF, funding program Photonics Research Germany ("LPI-BT1-FSU", FKZ 13N15466) and integrated into the Leibniz Center for Photonics in Infection Research (LPI). The LPI initiated by Leibniz-IPHT, Leibniz-HKI, UKJ and FSU Jena is part of the BMBF national roadmap for research infrastructures

Funding: IPHT Innovation Project 3D-HiRes 2021/2022 (Project No.: K690082). DFG via SFB 1375 NOA (Project No.: 398816777). DAAD-PPP-Japan Tsukuba University (project No.: 57710865). IRTG 2675 Meta-Active (Project No.: 437527638). DFG via CRC-TRR 234 CataLight (Project No. 364549901). DFG under Germany's Excellence Strategy, EXC 2051 (Project No. 390713860). EFRE Programm and Freistaat Thüringen, projects EU-0V/2020-59 (2019 FGI0017) and EU-0V/2023-1 (2022 FGI 0004). Photonics Research Germany ("LPI-BT1-FSU", FKZ 13N15466).

## Competing interests

The authors declare no competing interests.

## Data Availability Statement

The data that support the findings of this study are available from the corresponding author upon reasonable request.

Supplementary Information

# Two-Photon-Induced Direct 3D Printing of Freeform High-Index Phase-Change $Sb_2S_3$ Nanostructures

*Abhrodeep Dey[1,2], Andrea Dellith[1], Anne Sauer[3], Uwe Hübner[1], Henrik Schneidewind[1], Markus A Schmidt[1,4,5], Astrid Bingel[3], Volker Deckert[1,2,4,6], Jer-Shing Huang[1,2,4,7,8*], Wei Wang[1,2*]*

E-Mail: jershing.huang@ipht-jena.de; w.wang@uni-jena.de

[1] Leibniz Institute of Photonic Technology (IPHT), 07745 Jena, Germany.

[2] Institute of Physical Chemistry, Friedrich Schiller University Jena (FSU), 07743 Jena, Germany.

[3] Functional Optical Surfaces and Coatings, Fraunhofer IOF, 07745 Jena, Germany

[4] Abbe Center of Photonics and Faculty of Physics, FSU, 07745 Jena, Germany.

[5] Otto Schott Institute of Material Research, FSU, 07745 Jena, Germany.

[6] Jena Center for Soft Matter (JCSM), FSU, 07743 Jena, Germany.

[7] Research Center for Applied Sciences, Academia Sinica, 11529 Taipei, Taiwan.

[8] Department of Electrophysics, National Yang Ming Chiao Tung University, 30010 Hsinchu, Taiwan.

## S1: Comparison of different fabrication methods using Sb-BDCA based solution phase fabrication approach

| | g-EBL[1] | Positive-tone wet-etching[2] | DITPS (This work) |
|---|---|---|---|
| **Tone** | Negative tone (additive) | Positive tone (subtractive) | Negative tone (additive) |
| **Fabrication technique / Resolution** | EBL / resolution: sub-50 nm | fs-laser lithography / lateral resolution: 178 nm | Two photon solidification / lateral resolution: ~280 nm |
| **Depth control** | Limited 2.5D/Depth Control However, 2D/quasi 3D structures with high resolution can be fabricated. | 2.5D lithography / Depth control / selective material removal can be achieved to fabricate trenches, variable depth features and grooves. | Full freeform 3D / planar 2D control<br><br>2.5D control might be possible by controlling the voxel aspect-ratio via truncated voxels.[3] |
| **Substrates** | Only dielectric substrate. | Compatible on both metal and dielectric substrates. | Compatibility to conductive substrates. |
| **Unique advantages** | Direct patterning only on conventional substrates with a three-step approach. | Patterning on unconventional substrates (fiber end facets, camera sensors, reflective metals) can be achieved in a four-step approach. | Printing on unconventional substrates is possible in a three-step maskless approach.[4, 5] |
| **Throughput** | Point by point scan of EBL leads to low throughput. Makes it unsuitable for high-volume, large-area manufacturing.[6] | Voxel-based volumetric fabrication facilitates large scale patterning. | Large scale rapid prototyping can be achieved due to voxel-based volumetric printing. |

**Table S1:** Comparison between the g-EBL, positive tone and negative tone laser direct writing of $\alpha$-$Sb_2S_3$ structures using Sb-BDCA solution phase approach.

## S2: Miscellaneous DITPS printed freeform 3D structures

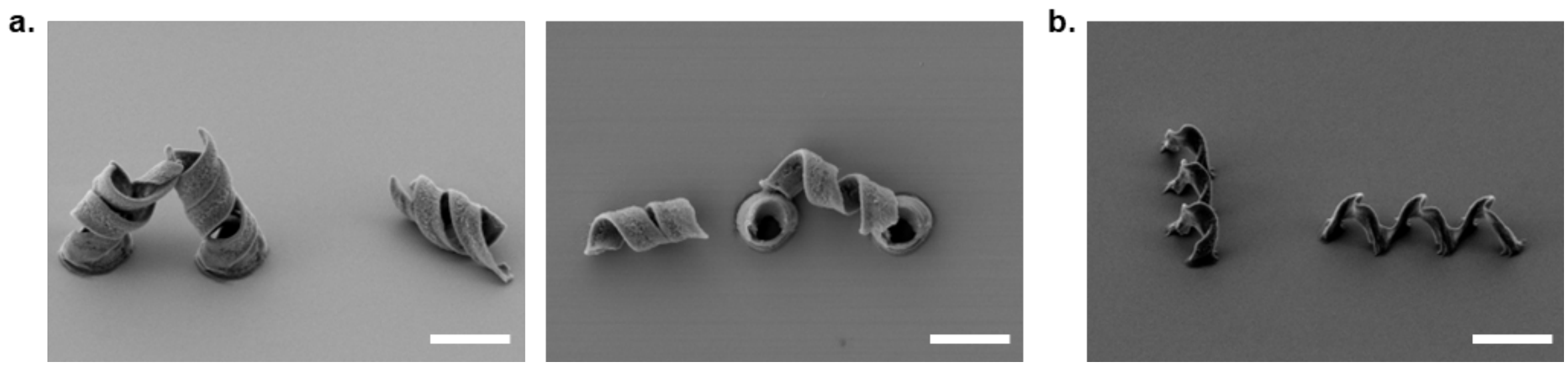

**Figure S2.** Short gallery of 3D printed structures on Au- and ITO-coated $SiO_2$ substrates. **(a)** (left) Double helical structures and (right) originally upright single helix structures on 50 nm Au-coated $SiO_2$ substrate **(b)** Single helix structures traversing parallel along the substrate plane on ITO-coated $SiO_2$ substrate. Scale bar: 10 µm.

## S3: EDX spectrum of DITPS structures

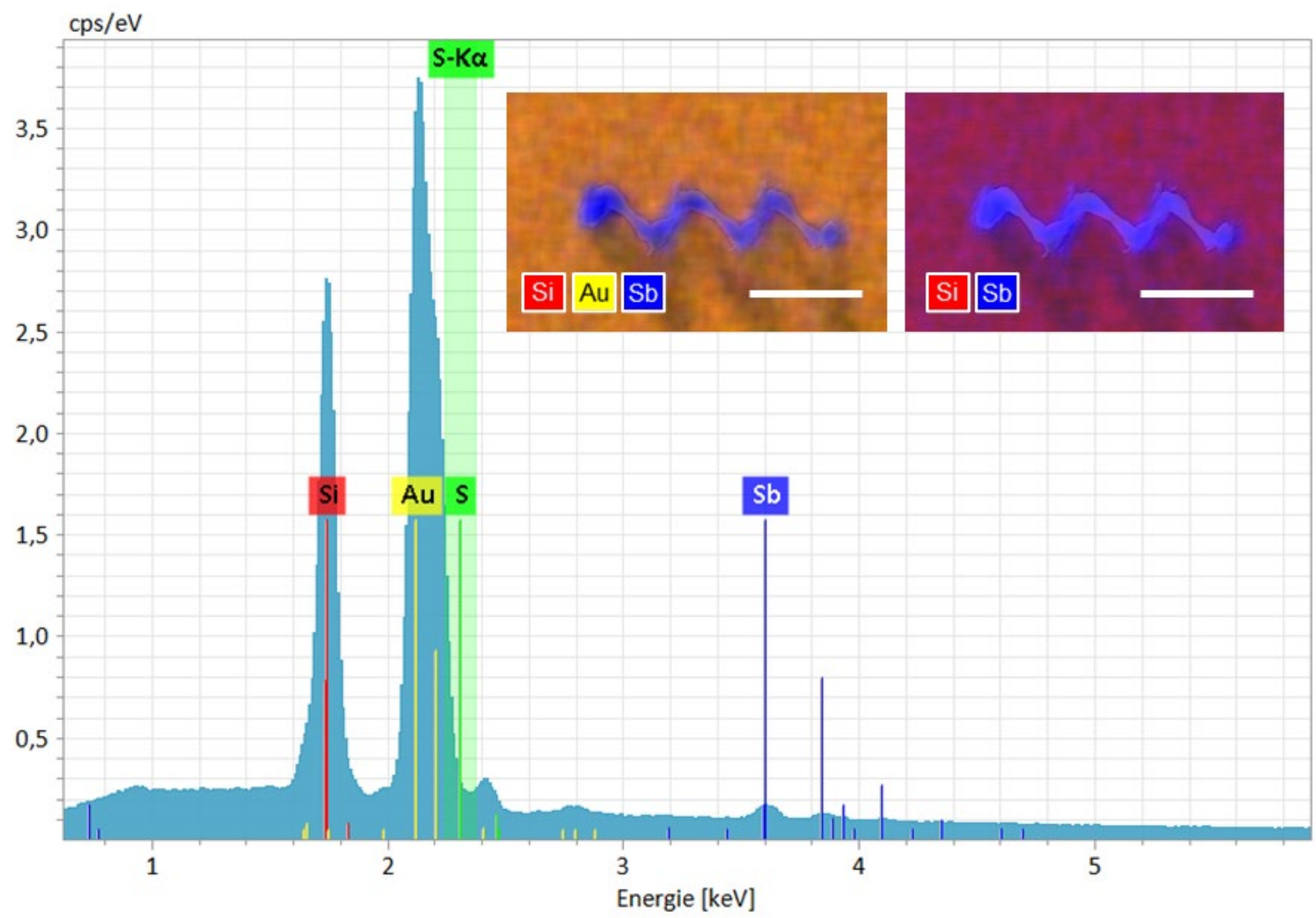

**Figure S3.1.** The EDX spectrum, obtained as an integral over the entire analyzed structure area, reveals a pronounced spectral overlap between the S-Kα and Au-Mα lines. In addition, the inherently weak S-Kα emission at 2.3 keV experiences significant attenuation due to the Au absorption edge near 2.2 keV. As a result, the presence of S cannot be reliably detected or visualized in this elemental mapping. Scale bar: 10 µm.

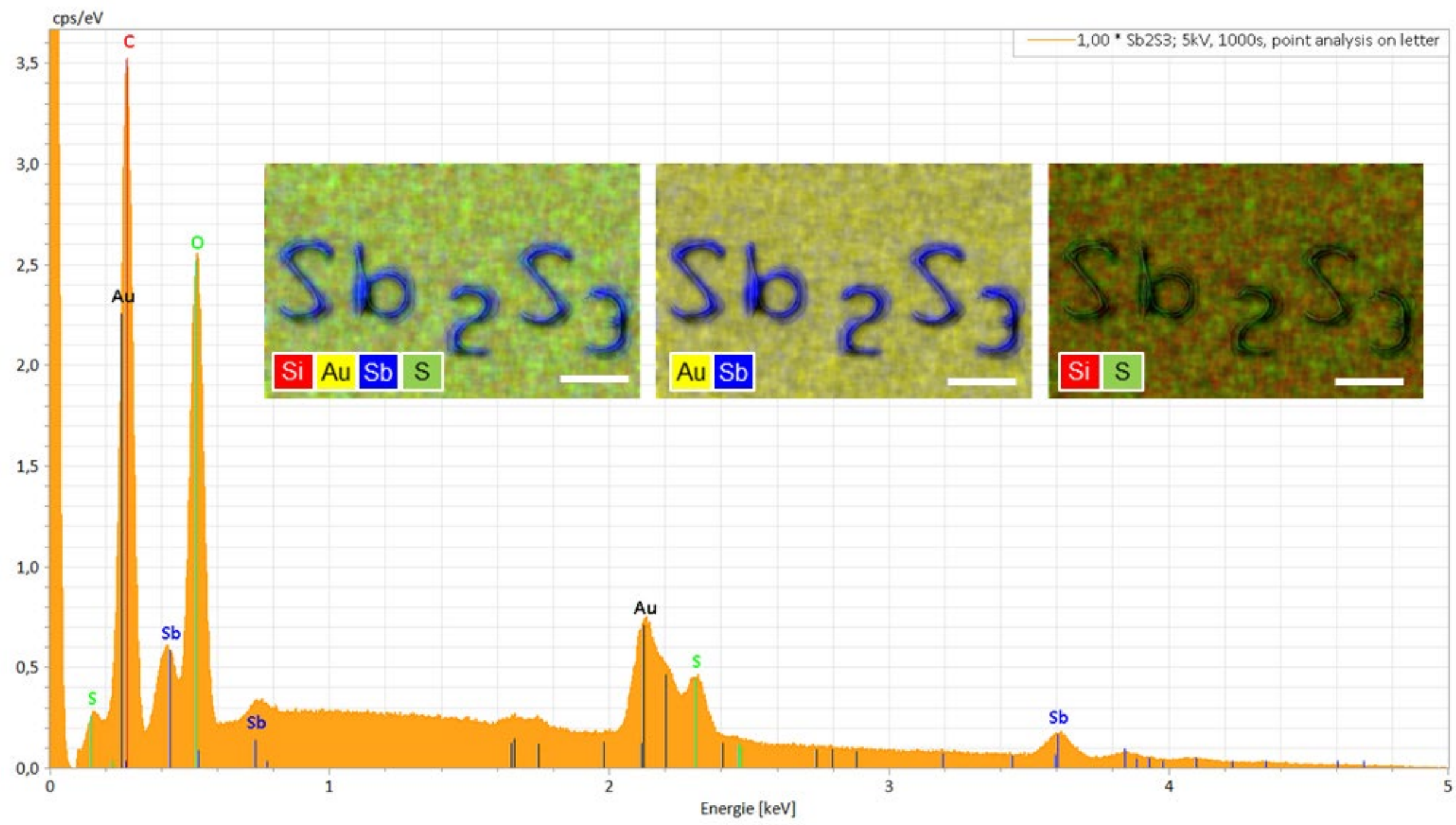

**Figure S3.2.** EDX spectrum acquired directly from the letter-shaped region of the Au-substrate sample confirms the presence of S in the material, as the contribution of gold from the substrate and the associated interference with S detection are significantly reduced in this measurement. Scale bar: 10 µm.